\documentclass{aa}
\usepackage{graphicx,natbib,amssymb}
\newcommand{\Msun}{\mbox{\rm M$_{\odot}$}}
\begin{document}
   \title{New nearby stars selected in a high proper motion survey by DENIS
photometry}

   \author{C\'eline Reyl\'e \inst{1} \and Annie C. Robin \inst{1} \and 
Ralf-Dieter Scholz \inst{2} \and Mike Irwin \inst{3}}

   \institute{CNRS UMR6091, Observatoire de Besan\c{c}on, BP1615, 
       25010 Besan\c{c}on Cedex, France \and
    Astrophysikalisches Institut Potsdam, An der Sternwarte 16, Potsdam,
       D-14482, Germany \and
    Institute of Astronomy, University of Cambridge, Madingley Road, 
       Cambridge, CB3 0HA, England\\
   email: celine@obs-besancon.f, annie.robin@obs-besancon.f, rdscholz@aip.de, mike@ast.cam.ac.uk\\
	}

   \offprints{C\'eline Reyl\'e}
   \date{Received ; accepted }


   \abstract{We present new nearby stars extracted from a proper motion catalogue and having a DENIS
counterpart. Their distances and spectral type are estimated using the DENIS colours. 107 stars are 
within 50~pc. 31 stars among them have previously measured distances. In addition, 40 
stars may enter within the 50~pc limit depending on which 
population they belong to. 6 stars among them have already measured distances. 5 objects, LHS5045, 
LP225-57, LP831-45, 
LHS1767, and WT792, are probably closer than 15~pc, with LP225-57 at 9.5~pc. Most of these stars 
are M-type while 4 stars are white dwarfs. 88 M-dwarfs are disc stars, 14 belong to the thick disc
and 1 to the spheroid.
   }

   \maketitle

\section{Introduction}

The solar neighbourhood serves as a fundamental constrain for our understanding of the stellar 
physics and the Galaxy. The nearest stars provide accurate data on luminosities, 
temperatures, masses, which are fundamental parameters of stellar astronomy. Their
kinematics, chemical composition, and mass function also hold important clues to the nature and 
history of the Milky Way.

However, among the nearest stars, many of the low luminosity stars such as white, red, and brown
dwarfs remain undetected. \cite{henry1997} estimated that more than 30~\% of the stars within 10~pc 
are currently missing from the solar neighbourhood sample. The deficit is estimated to be twice 
within 25~pc \citep{henry2002} and is largest south of decination -30$^\circ$. As a confirmation,  
stars are continuously identified as nearby stars in proper motion catalogues or in Schmidt
plates \citep{henry1997,henry2002,gizis1997,scholz1999,scholz2001,scholz2002,jahreiss2001,phanbao2001,delfosse2001}.

In addition, new nearby faint stars, mainly brown dwarfs, are detected in deep sky surveys
\citep{ruiz1997,delfosse1997,delfosse1999,martin1999,kirkpatrick1999,kirkpatrick2000,reid2000,fan2000,gizis2000}.

One way to detect nearby stars is to select the red faint objects obtained in near-infrared surveys: 
the \textit{Two Micron All Sky Survey} (2MASS) \citep{kirkpatrick1999,kirkpatrick2000,reid2000,gizis2000} 
and the \textit{Deep Near-Infrared Survey of the 
Southern Sky} (DENIS) \citep{delfosse2001,phanbao2001}. Most of the nearest stars found in these surveys
turned out to be high proper motion objects. Given this fact, \cite{scholz2001} searched for new
stars in the solar neighbourhood by combining proper motion catalogues with the 2MASS data base.

Using the same approach, we cross-identify a proper motion catalogue with DENIS data 
and estimate the distances using near-infrared photometry.
The proper motion catalogue is described in Sect.~\ref{catalogue}. In Sect.~\ref{Xid}, we give the 
result of the cross-identification with the DENIS data base. Sect.~\ref{dist} presents the distance
estimation. In Sect.~\ref{feh}, we discuss
the effect of metallicity on the distance estimation. In Sect.~\ref{results}, we list the stars found
to be within 50~pc and discuss the precision on our distance determinations. 

\section{High proper motion sample}
\label{catalogue}

One of the criteria for finding nearby stars is their large proper motion. The main source
for the identification of such stars are the high-proper motion catalogues of Luyten: the \textit{Luyten 
Half Second proper motion catalogue} (LHS) and the \textit{New Luyten Catalogue of Stars with Proper 
Motions Larger than Two Tenths of an Arcsecond} (NLTT) based on observations with the Palomar Schmidt 
telescope. \cite{dawson1986} showed that the LHS catalogue is incomplete south of declination 
-33$^\circ$, $R > 18$ or $\mu > 2.5~''$~yr$^{-1}$. 
\cite{wroblewski1989,wroblewski1991,wroblewski1994,wroblewski1996,wroblewski1997} improved the
completness in the southern sky and found new stars with $\mu > 0.15~''$~yr$^{-1}$ and photographic 
magnitudes $m_{pg} < 20$. \cite{ruiz1993} found proper motion stars in the ESO Schmidt plates with 
$\mu > 0.1~''$~yr$^{-1}$ and $R < 20$.

In order to complete the existing proper motion catalogues at fainter magnitudes in the southern
hemisphere, \cite{scholz2000} used Automatic Plate Measuring (APM) measurements of sky survey plates 
taken with the UK Schmidt 
telescope. Plates in the passbands $B_J$ and $R$ were taken with typical epoch difference of 
15~years.
The resulting catalogue contains 693 stars (out of which 195
have already been published by \cite{scholz2000}) south of declination -15$^\circ$, in the regions 
0$^h$-7$^h$, 10$^h$-14$^h$, and 19$^h$-23$^h$ in right ascension. 
The overall search area is about 3000 square degrees. The lower proper motion
limit of the survey is typically 0.25 arcsec yr$^{-1}$. But in fields with smaller 
epoch difference between the $B_J$ and $R$ plates, only objects with even 
larger proper motions could be detected, respectively. The distribution of 
the epoch differences as a function of declination was shown in \cite{scholz2000}. 
The largest proper motions found are about 1.3 arcsec yr$^{-1}$.  The 
practical limit for searching for high proper motion objects using two plates 
in different passbands is about 1 magnitude above the respective plate limits, 
i.e. $R = 20.0$ and $B_J = 21.5$. For a comparison with other high proper 
motion catalogues and an estimate of the completeness of the survey, see 
\cite{scholz2000}. About half of the objects detected are known proper
motion stars,
whereas the remaining objects mainly at fainter magnitudes are new high proper 
motion stars. 

With these measurements, they were able to fill the
gap in the high proper motion survey of faint stars with $\delta <$ -33$^\circ$. In addition they
obtained more accurate positions and photometry for the already known  high proper motion
stars. The accuracies are 0.1 to 0.2~$''$ in position, 0.03~$''$~yr$^{-1}$ on proper motion, and 
$\sim$~0.2~mag on photometry.

\section{Cross-identification with DENIS}
\label{Xid}

The DENIS survey \citep{epchtein1997} provides a full coverage of the southern sky in the
optical band $I$ (0.85~$\mu$m) and the near-infrared bands $J$ (1.25~$\mu$m) and $K_s$ (2.17~$\mu$m).
The position accuracy is 0.5~$''$ and the photometric accuracy is better than 0.1 mag. About 40\% of the 
data are up to now calibrated at the \textit{Paris Data Analysis Center} (PDAC). 

\begin{figure}[h]
\centering
\includegraphics[width=7.4cm,clip=,angle=-90]{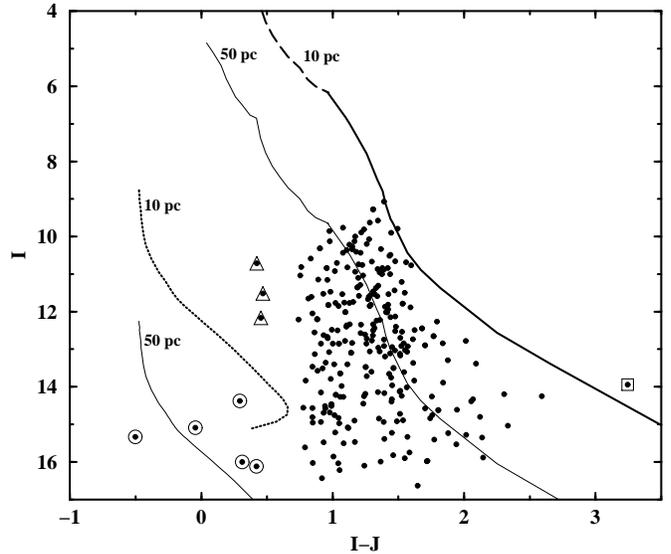}
   \caption{($I$,$I-J$) colour-magnitude diagram for the high proper motion stars cross-identified with 
DENIS. Dots: G, K, or M-dwarfs. Square: possible brown dwarf. Circles: white dwarfs. Triangles: G-subdwarfs.
Solid line: theoretical relation for M-dwarfs with solar metallicity at 10~pc \citep{baraffe1998}.
Dashed line: theoretical relation for G and K-dwarfs with solar metallicity at 10~pc \citep{lejeune1997}.
Dotted line: theoretical relation for white dwarfs with 0.6~\Msun~at 10~pc \citep{bergeron1995}. The thin
lines show the same relations for stars at 50~pc.}
   \label{fig1}
\end{figure}

We look for counterparts of the given proper motion sample in the DENIS point source 
catalogue extracted at the PDAC with a search radius of $3''$.
We get 301 recoveries among the 693 objects. None of them have a counterpart in the USNO A2.0
catalogue, as expected for proper motion stars. The density of stars with no USNO counterpart in a 
field in the Galactic plane is less than 0.01 star within the search region. Thus the possibility
of false cross-identification is low, smaller than 1\%.
Most of the unrecovered stars are in the unprocessed part of the DENIS survey.
Dots in Fig.~\ref{fig1} shows the ($I$,$I-J$) colour-magnitude diagram of the recovered stars. 
Most stars have $I-J >$~1. They are M-dwarfs \citep{leggett1992}. Stars with $0.5~<~I-J~<~1$ are
G or K-dwarfs. 1 star, with $I-J~>$~3, is a possible brown dwarf (square). Stars with $I-J~<$~0.5 are 
either white dwarfs (circles) or distant subdwarfs (triangles), as discussed in Sect.~\ref{dist}. 

\section{Photometric distance estimation}
\label{dist}

Photometric distances are estimated using the $I-J$ colour index. Theoretical colour-magnitude relations ($M_I$,$I-J$) for stars placed at 10~pc are superimposed 
to the observations in 
Fig~\ref{fig1}. The solid line shows the relation for M-dwarfs with solar metallicity 
\citep{baraffe1998}, the dashed line shows the relation for solar metallicity G and K-dwarfs 
\citep{lejeune1997}, and the dotted line shows the relation for white dwarfs with 0.6~\Msun 
\citep{bergeron1995}. 

We compute photometric distances from the DENIS $I-J$ colour index and the corresponding theoretical 
colour-magnitude relation: \cite{baraffe1998} relation for M-dwarfs with $I-J >$ 1, \cite{lejeune1997}
relation for G and K-dwarfs in the range 0.5 $< I-J <$ 1. Stars with $I-J$ $<$ 0.5 can be either 
nearby white dwarfs or distant subdwarfs. The 5 faintest blue stars (circles in Fig.~\ref{fig1}) are 
white dwarfs. Indeed if they would be subdwarfs with that faint apparent magnitude, they would be at 
$d >$ 1600 pc. Their proper motion would
lead to a tangential velocity $>$ 4000 km s$^{-1}$, much larger than the Galactic escape speed 
\citep{leonard1990,meillon1997}. For these stars, we use \cite{bergeron1995} model atmosphere for 
white dwarfs of mass 0.6 \Msun, before the turnover ($M_I \lesssim$ 14.5), assuming they are not 
cool white dwarfs.

The 3 brighter blue stars (triangles in Fig.~\ref{fig1}) are more probably late G-stars of the 
spheroid with large tangential velocity, rather than disc white dwarfs closer than 3~pc. The theoretical 
relation from \cite{lejeune1997} for the typical spheroid metallicity $[\frac{Fe}{H}] = -1.8$ gives 
$M_I \sim$ 5~mag at $I-J \sim$ 0.5. The derived photometric distances of these 3 stars ranges between 
160 and 260~pc, and their tangential velocities between 280 and 380~km~s$^{-1}$, in agreement with the
spheroid kinematics.

$I-K_s$ is also a good estimator of the
luminosity or effective temperature of M-dwarfs. Photometric distances differ by 15\% depending on which 
colour index is used. However, 
\cite{delfosse1997} showed that the $K_s$ band of DENIS can differ by $\sim$0.1~mag compared with the
standard $K$ band for which models are computed. We then choose to estimate the distances from the 
$I-J$ colour index and ($M_I$,$I-J$) theoretical relations. 

\section{Metallicity effect}
\label{feh}
Within the sample magnitude range and proper motion range, one expects to get disc stars, but also 
thick disc and spheroid stars. Using a theoretical colour-magnitude relation at solar metallicity 
would overestimate the distance for the stars of these old populations.
We use the Besan\c{c}on model of population synthesis to reproduce the stellar 
content in the fields of the sample. The Besan\c{c}on model has been described in \citep{haywood1997}.
Since then, \cite{Robin2000} improved the halo density law and mass function, while \cite{Reyle2001} 
constrained the thick disc population description. New constraints available on the galactic
potential, the local luminosity function and the kinematics versus age, obtained with the
Hipparcos mission, are also taken into account to derive the thin disk density law (Reyl\'e et al.,
to be submitted).

\begin{figure}[h]
\centering
\includegraphics[width=7.4cm,clip=,angle=-90]{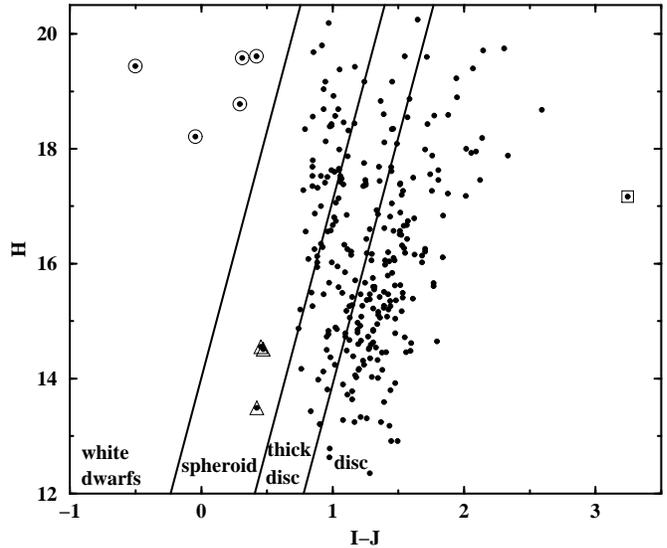}
   \caption{($H$,$I-J$) diagram for the high proper motion stars cross-identified with 
DENIS. Symbols have the same meaning as in Fig.~\ref{fig1}. The solid lines define the regions 
where the stars are most probably, from right to left: disc stars, thick disc stars, spheroid 
stars, white dwarfs.}
   \label{fig2}
\end{figure}

The simulation shows that the different
populations are separated in the plane ($H, I-J$), where $H = I + 5 \mbox{log}\mu +5$ is the reduced
proper motion. 
We define three regions in this plane where stars are most probably disc stars, thick 
disc stars, or spheroid stars, as represented in Fig.~\ref{fig2}. 

We determine the photometric distance using a theoretical colour-magnitude relations ($M_I$,$I-J$) 
at solar metallicity for stars in the disc region, at metallicity $[\frac{Fe}{H}] = -0.8$ 
in the thick disc region, and  at metallicity $[\frac{Fe}{H}] = -1.8$ in the spheroid region. 
These relations are obtained by interpolation of the available relations with 
$[\frac{Fe}{H}] = -0.5, 
-1, -1.5, -2$ from \cite{lejeune1997} for G and K-stars and \cite{baraffe1998} (private 
communication for low metallicities) for M-stars.

For stars in between two regions, we make two determinations with both metallicity hypotheses
(disc/thick disc or thick disc/spheroid). The probability that a star with a magnitude $I$,
a colour index $I-J$ and a proper motion $\mu_{tot}$, belongs to one population 
is estimated from the population distribution of the stars in the simulated sample. In this case, 
both distances and tangential velocities are computed. 

Disc stars with tangential velocities as large as 250~km~s$^{-1}$ still are found with a large 
probability in the simulated sample. Indeed, the sample is not representative of the entire disc 
population, but as a high proper motion sample, it selects the stars having a high velocity.

\section{Stars within 50~pc}
\label{results}
107 stars are found to have photometric distances less than 50~pc. They are mainly M-stars, but 4 stars 
are white dwarfs. 
According to the colour-reduced proper motion diagram ($H, I-J$), 88 stars among them belong 
to the disc, 14 to the thick disc stars, and 1 to the spheroid. In addition, 33 stars belong either
to the disc or to the thick disc. Distance estimations with both hypothesis are made for 
these stars and they are possibly within 50~pc. Likewise, 7 stars belong either to the
spheroid or to the thick disc and may enter or not within the 50~pc limit.

Astrometric, photometric and 
kinematical characteristics, estimated distances and spectral types are given in Table~1. The spectral 
types for M-stars are derived from the $I-J$ colour index \citep{leggett1992}. 4 late M-type stars have 
been previously identified by \cite{phanbao2001} and 1 by \cite{henry2002}. 6 stars already have 
spectroscopic distance estimates \citep{scholz2002}. The extremely red object 
is the already known brown dwarf LP944-20 at 5 pc \citep{tinney1996}. Trigonometric distances from
the Yale parallax catalogue or the Hipparcos catalogue, or photometric distances from the ARI 
database for nearby stars (ARICNS), available for 25 stars, are also indicated in Table~1.

Among the list of 107 nearby stars, 15 new stars are within the
25~pc limit of the \textit{Catalogue of Nearby Stars} \citep[CNS3,][]{gliese1991}.
New discoveries within 15~pc are LHS5045, LP225-57, LP831-45, LHS1767, and WT792.
LP225-57 is presumably closer than 10~pc, at $d$ = 9.5 pc. Its estimated spectral type 
is M3.5. The star LHS124 may be a M1-star of the thick disc at 7.2~pc, although the
simulation of the stellar sample tends to show it is more likely a more distant star in the disc.
The distance indicated in the ARICNS (21~pc) seems also to favor a disc star.

It should be noted that the distances estimated in Table~1 have large uncertainties, particularly
for stars with $I-J <$ 1.5, for which the colour index is less sensitive to the luminosity (see
Fig.~\ref{fig1}). Moreover, \cite{phanbao2001} plotted the colour-magnitude relation of M-stars with 
measured distances and noted an intrinsic scatter of $\pm$~1~mag. They also superimposed the 
\cite{baraffe1998} theoretical track which does not fit perfectly the mean observed relation around 
$I-J$ = 1.5 where the $M_I$ magnitude is lower by 1 magnitude compared to the magnitude obtained with
a polynomial fit of the observed data \citep{phanbao2001}. For this reason, our distance determination
for the 4 stars already identified by \cite{phanbao2001} is smaller by a factor up to 15\%.

\begin{figure}[h]
\centering
\includegraphics[width=7.4cm,clip=,angle=-90]{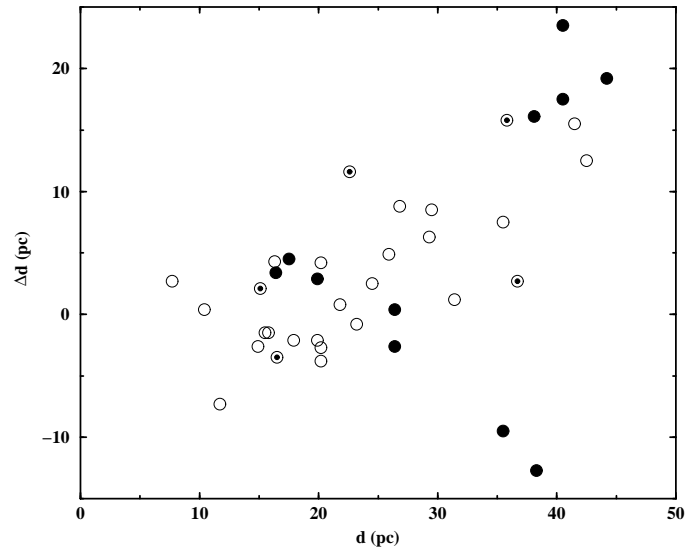}
   \caption{Comparision between our estimated distances and determination by other authors, as
indicated in Table~1. $\Delta d$ is the difference between both determinations. $d$ is our
determination. Open circles: photometric determinations. Filled circles: trigonometric
determinations. Circled dots: spectroscopic determinations.}
   \label{fig3}
\end{figure}

Fig.~\ref{fig3} gives the difference between our estimated distance and other distances found in the
litterature, as indicated in Table~1 footnotes, versus our determination. Open circles are used 
when the other determination is photometric, filled circles when it is trigonometric, circled dots
when it is spectroscopic.
Our determination is not biased compared to trigonometric parallaxes, giving confidence in our
distances, while a systematic deviation appears at large distances between other photometric distances 
and ours, due to the different theoretical colour-magnitude relations used. The mean error is of the 
order of 5~pc at distances closer than 25~pc but grows to more than 10~pc at larger distances.

\begin{table*}
\label{candidats}
\rotatebox{90}{
{\scriptsize 
\begin{tabular}{llcccccrccrrrrlll}
\multicolumn{17}{p{24cm}}{{\bf \normalsize Table 1.} \normalsize Nearby stars candidates. The distance $d$ 
is estimated from the $I-J$ colour index. When two distances are indicated, they are obtained with two 
metallicity hypothesis and their probable population with probability are indicated in the last two 
columns (see text): I = disc, int = thick disc, II = spheroid. The tangential velocity is also 
computed with both distance estimates. The spectral 
types for M-stars are derived from \cite{leggett1992}. wd means white dwarf.}\vspace*{0.3cm}\\
\hline
name	&other  &$\alpha_{J2000}$&$\delta_{J2000}$&UKST		&$R$	&$B_J-R$ &\multicolumn{1}{c}{$I$}	&$I-J$	&$J-K_s$	
&\multicolumn{1}{c}{$d$}	&\multicolumn{1}{c}{$\mu_x$}	&\multicolumn{1}{c}{$\mu_y$}	
&\multicolumn{1}{c}{$V_t$}	&type &\multicolumn{2}{l}{population \%} \\	
APMPM &name	&		&		&epoch		&UKST	&UKST	&DENIS	&DENIS	&DENIS	&\multicolumn{1}{c}{(pc)}
&\multicolumn{2}{c}{($''$ yr$^{-1}$)}	&(km s$^{-1}$)	& && \\
\hline
J0001-3650	&LP987-71 	&00 00 34.09 	&$-$36 50 09.3  &1996.61 	&13.28 	&1.75 	&12.78  &1.09 	&0.82 	&43.9 	    &0.40    &0.11    &85.3 	   &M0.5 &int	&\\
J0003-3414$^1$	&LHS1008	&00 02 39.95 	&$-$34 13 37.7  &1996.61   	&14.22  &0.45   &14.38	&0.29	&0.63  	&17.5	    &0.15    &$-$0.74 &63.0        &wd   &	&\\
J0009-2707$^2$	&LHS1026	&00 09 03.88	&$-$27 07 22.2	&1990.80	&10.08	&1.92	&9.76	&1.08  	&0.64	&40.5/11.6  &0.66    &0.11    &128.6/36.8  &M0.5 &I/int 	&85/15\\
J0013-3757	&WT3		&00 13 29.87	&$-$37 56 44.8	&1989.66	&13.24  &2.29	&12.85	&1.00	&0.78	&73.8/33.7  &0.07    &$-$0.58 &204.4/93.3  &M0   &int/II&85/15\\
J0015-5147	&LP218-29	&00 15 15.02	&$-$51 47 04.8  &1993.69 	&11.04 	&2.66 	&11.15  &0.96   &0.84 	&100.0/42.6 &0.33    &0.09    &161.2/68.7  &M0   &I/int 	&35/65\\
J0017-4149	&WT4		&00 16 34.85 	&$-$41 49 19.1	&1989.66	&14.80  &2.14	&13.98	&1.07 	&0.89	&80.5/44.0  &0.37    &$-$0.33 &189.2/103.4 &M0.5 &int/II&50/50\\
J0017-5016$^3$	&LP218-9	&00 16 36.18 	&$-$50 16 10.9 	&1993.69 	&10.12 	&2.48 	&10.12  &1.16 	&0.88 	&38.1 	    &0.31    &0.29    &75.8	   &M1   &I	&\\
J0020-5743	&LP170-67	&00 20 15.88 	&$-$57 42 40.9 	&1990.78 	&13.11 	&2.26 	&12.02  &1.41  	&0.80  	&35.8 	    &0.39    &0.12    &69.6        &M3   &I	&\\
J0023-4833	&		&00 22 34.65 	&$-$48 32 35.6  &1993.69 	&13.72 	&2.19 	&12.83  &1.15  	&0.97 	&138.8/35.2 &0.31    &0.10    &217.1/55.1  &M0.5 &I/int	&35/65\\
J0033-4733	&LP291-115	&00 33 13.26	&$-$47 33 17.8	&1993.69	&14.20	&2.25	&12.26	&1.79  	&0.94	&15.7	    &0.26    &0.15    &22.3	   &M5   &I	&\\
J0034-3759	&LP937- 57	&00 33 42.79	&$-$37 59 10.5	&1989.66	&11.74	&1.84	&11.04	&1.23	&0.74	&48.3	    &0.18    &0.41    &103.0	   &M1   &I	&\\
J0049-6102$^4$	&LHS124		&00 49 27.58 	&$-$61 02 32.4 	&1990.78 	&10.52 	&1.51 	&9.89 	&1.22 	&0.76 	&29.5/7.2   &1.09    &$-$0.04 &152.4/37.2  &M1   &I/int	&75/25\\
J0058-2751$^5$	&LHS129		&00 58 26.55 	&$-$27 51 21.7 	&1986.77 	&10.28 	&1.87 	&9.28 	&1.31 	&1.14 	&16.4 	    &1.26    &$-$0.32 &101.1	   &M2.5 &I	&\\
J0059-3127	&LHS1171	&00 58 44.47	&$-$31 27 09.0	&1986.77	&12.93	&2.24	&11.52	&1.31	&0.75	&44.7	    &0.50    &$-$0.40 &135.6	   &M2.5 &I	&\\
J0102-6739	&LHS1178	&01 02 05.88	&$-$67 39 20.2	&1987.82	&13.84  &1.80	&12.85	&1.05	&0.75	&55.2/28.4  &0.88    &$-$0.18 &235.0/120.9 &M0   &int/II&50/50\\
J0127-3219	&LHS1251	&01 27 02.57	&$-$32 18 50.2  &1991.91 	&12.28 	&2.09 	&11.85	&1.04 	&0.81 	&36.0 	    &0.20    &$-$0.52 &95.6 	   &M0   &int	&\\
J0149-6427	&LP88-1		&01 48 32.97	&$-$64 26 44.0 	&1987.90 	&12.02 	&1.82 	&10.92	&1.02 	&0.86 	&79.3/26.7  &0.41    &0.20    &172.9/58.2  &M0   &I/int	&30/70\\
J0150-3319	&LP940-20	&01 49 42.37	&$-$33 19 21.5	&1991.91	&14.36	&2.24	&12.65	&1.77	&0.88	&19.6	    &0.38    &0.11    &37.2	   &M5   &I	&\\
J0151-3058	&LP884-94	&01 51 04.98	&$-$30 57 59.0	&1994.89	&14.02	&2.60	&12.52	&1.47	&...	&35.5	    &0.08    &0.31    &53.8	   &M3.5 &I	&\\	
J0153-4806	&LHS5045	&01 52 52.07 	&$-$48 05 39.2 	&1988.91 	&12.40 	&2.12 	&10.77 	&1.60 	&0.91 	&11.0 	    &$-$0.55 &$-$0.20 &30.8 	   &M4   &I	&\\
J0155-5307	&LP223-77	&01 55 12.82 	&$-$53 06 34.0 	&1990.93 	&13.65 	&2.54 	&12.13 	&1.48 	&0.88 	&29.6 	    &0.26    &0.22    &47.7	   &M3.5 &I	&\\
J0158-5031	&		&01 57 54.88 	&$-$50 30 56.1  &1988.91   	&17.29  &2.23  	&14.61	&1.95   &...    &37.8	    &0.71    &$-$0.10 &129.0	   &M5.5 &I	&\\
J0211-6314$^6$	&LHS1351	&02 11 19.04 	&$-$63 13 37.6 	&1989.73 	&10.77 	&2.08 	&9.80 	&1.24 	&0.81 	&26.8 	    &$-$0.70 &$-$0.31 &97.8	   &M1   &I	&\\
J0211-4714	&		&02 11 26.70 	&$-$47 13 40.3  &1988.91 	&13.75 	&1.83 	&13.09	&1.14 	&0.80 	&41.3       &0.35    &0.23    &82.2  	   &M0.5 &int	&\\
J0213-3352	&LHS1355	&02 12 39.48	&$-$33 52 07.6	&1996.62	&14.21	&2.12	&12.70	&1.52	&0.93	&33.9	    &0.83    &0.25    &139.8	   &M3.5 &I	&\\	
J0213-7346$^7$	&LHS1356	&02 12 58.07 	&$-$73 45 51.9 	&1993.85 	&11.24 	&1.44 	&10.40 	&1.18 	&0.90 	&41.5 	    &0.47    &0.25    &104.3	   &M1   &I	&\\
J0218-5508	&LP174-34	&02 18 17.41 	&$-$55 07 34.6 	&1990.93 	&12.25 	&1.69 	&10.87 	&1.34 	&1.00 	&30.0 	    &0.45    &0.20    &69.7	   &M2.5 &I	&\\
J0220-6519$^8$	&		&02 19 55.74 	&$-$65 18 44.8 	&1989.73 	&14.93 	&2.02 	&12.92 	&1.52 	&0.86 	&36.7 	    &0.51    &0.11    &90.5	   &M3.5 &I	&\\
J0224-4547	&		&02 24 17.19 	&$-$45 46 55.8 	&1990.73 	&14.05 	&1.63 	&12.57	&1.21 	&0.87 	&102.9/25.3 &0.33    &0.19    &185.3/45.6  &M1   &I/int	&75/25\\
J0228-8056	&		&02 27 54.63 	&$-$80 55 37.6 	&1991.63 	&14.78 	&1.81 	&12.94 	&1.45 	&0.94 	&46.2 	    &$-$0.17 &$-$0.34 &83.2 	   &M3   &I	&\\
J0232-2725	&LHS5057	&02 32 28.72 	&$-$27 25 13.4  &1994.93 	&17.00 	&2.23 	&15.33	&1.57 	&0.82 	&96.3/45.1  &0.38    &0.23    &200.8/94.1  &M4   &I/int	&60/40\\
J0233-2737	&LHS1416	&02 32 41.86 	&$-$27 37 26.0 	&1994.93 	&14.06 	&2.60 	&13.06	&1.36 	&0.83 	&22.1       &0.65    &0.37    &78.6  	   &M3   &int	&\\
J0234-5306	&LP225-57	&02 34 20.95	&$-$53 05 35.4	&1993.96	&10.78	&2.10	&9.79	&1.49	&0.90	&9.6	    &0.25    &$-$0.34 &19.1	   &M3.5 &I	&\\
J0241-6045	&LP127-33	&02 40 36.99	&$-$60 44 48.5 	&1991.78 	&13.85 	&1.50 	&12.34 	&1.13 	&0.91 	&116.7/30.4 &0.19    &0.29    &193.6/50.4  &M0.5 &I/int	&45/55\\
J0244-5835	&		&02 43 56.21 	&$-$58 34 53.7 	&1991.78 	&16.56 	&2.28 	&13.89  &1.84 	&1.00 	&31.0 	    &0.29    &0.26    &57.3	   &M5   &I	&\\
J0244-5805	&LHS1437	&02 43 56.23 	&$-$58 04 51.9  &1991.78  	&13.35  &1.83   &12.70 	&0.96 	&0.80	&87.1/39.8  &$-$0.51 &$-$0.30 &244.3/111.7 &M0   &int/II&75/25\\
J0246-4025	&LP993-119	&02 45 59.15	&$-$40 24 55.8	&1993.96	&12.60	&1.70	&11.37	&1.33  	&0.92	&39.5	    &0.43    &0.02    &80.5	   &M2.5 &I	&\\
J0247-5257	&		&02 47 20.37 	&$-$52 56 48.2 	&1993.96 	&15.32 	&2.08 	&13.37	&1.61 	&0.98 	&35.5 	    &0.41    &0.53    &112.7	   &M4   &I	&\\
J0248-3043	&LHS1448	&02 48 19.52  	&$-$30 43 20.4 	&1994.93 	&10.64  &1.34 	&10.70 	&1.03 	&0.82 	&69.2/22.4  &0.38    &$-$0.56 &223.0/72.2  &M0   &I/int	&25/75\\
J0250-2444	&LP830-44	&02 49 36.12  	&$-$24 44 14.5 	&1992.68 	&13.90 	&1.49 	&12.40 	&1.10 	&0.88 	&128.8/35.1 &0.36    &0.03    &219.8/59.9  &M0.5 &I/int	&20/80\\
J0250-4941	&		&02 50 04.24  	&$-$49 40 42.4	&1992.81 	&15.68 	&2.29 	&14.17 	&1.39 	&0.87 	&108.1/34.3 &0.13    &0.28    &158.8/50.4  &M3   &I/int	&80/20\\
\hline\\
\multicolumn{14}{l}{$1$ distance is 13 pc from the Yale parallax catalogue}\\
\multicolumn{14}{l}{$2$ distance is 17 pc from the Yale parallax catalogue, 23 pc from the Hipparcos catalogue}\\
\multicolumn{14}{l}{$3$ distance is 22 pc from the Hipparcos catalogue}\\
\multicolumn{14}{l}{$4$ distance is 21 pc from the ARICNS}\\
\multicolumn{14}{l}{$5$ distance is 13 pc from the Yale parallax catalogue and the Hipparcos catalogue}\\
\multicolumn{14}{l}{$6$ distance is 18 pc from the ARICNS}\\
\multicolumn{14}{l}{$7$ distance is 26 pc from the ARICNS}\\
\multicolumn{14}{l}{$8$ already identified by \cite{scholz2002} at 34 pc}\\
\end{tabular}
}}
\end{table*}

\begin{table*}
\rotatebox{90}{
{\scriptsize 
\begin{tabular}{llcccccrccrrrrlll}
\multicolumn{14}{l}{{\bf \normalsize Table 1.} \normalsize Continued\vspace*{0.3cm}}\\
\hline
name	&other  &$\alpha_{J2000}$&$\delta_{J2000}$&UKST		&$R$	&$B_J-R$ &\multicolumn{1}{c}{$I$}	&$I-J$	&$J-K_s$	
&\multicolumn{1}{c}{$d$}	&\multicolumn{1}{c}{$\mu_x$}	&\multicolumn{1}{c}{$\mu_y$}	
&\multicolumn{1}{c}{$V_t$}	&type &\multicolumn{2}{l}{population \%} \\	
APMPM &name	&		&		&epoch		&UKST	&UKST	&DENIS	&DENIS	&DENIS	&\multicolumn{1}{c}{(pc)}
&\multicolumn{2}{c}{($''$ yr$^{-1}$)}	&(km s$^{-1}$)	& && \\
\hline
J0252-2602	&LHS1459	&02 51 43.22  	&$-$26 01 33.0 	&1992.68 	&13.21 	&1.57 	&11.76 	&1.09 	&0.82 	&28.0       &$-$0.01 &$-$0.82 &108.8  	   &M0.5 &int	&\\
J0252-3412	&LHS1461	&02 52 17.91  	&$-$34 11 51.0 	&1993.80	&14.38 	&1.73 	&12.77  &1.30   &0.77 	&84.2/21.9  &0.47    &0.09    &191.6/49.8  &M2.5 &I/int	&85/15\\
J0254-2837	&LP886-61	&02 54 08.29	&$-$28 36 35.8	&1994.93	&14.77	&2.14	&13.61	&1.54  	&0.88	&47.1	    &0.29    &0.17    &75.9	   &M4   &I	&\\
J0255-2216	&LP831-1	&02 54 39.26	&$-$22 15 58.0	&1992.68	&12.16	&1.69	&10.34	&1.37  	&0.69	&21.3	    &0.37    &$-$0.07 &38.4	   &M3   &I	&\\
J0314-2310	&LP831-45	&03 14 17.96	&$-$23 09 31.2	&1992.99	&11.46	&1.83	&9.90	&1.44  	&0.91	&11.7	    &0.35    &0.19    &22.2	   &M3   &I	&\\
J0319-3749	&LP943-70	&03 18 44.50	&$-$37 48 56.9	&1991.78	&15.15	&2.12	&13.02	&1.55   &0.80	&35.2	    &0.37    &$-$0.22 &71.7	   &M4   &I	&\\
J0319-4052	&LP994-96	&03 18 45.25	&$-$40 51 37.2	&1991.78	&10.64	&2.10	&9.63	&1.28  	&0.87	&21.2	    &0.16    &0.31    &35.2	   &M2.5 &I	&\\
J0329-2719$^9$	&LHS1549+1550	&03 28 48.15	&$-$27 19 05.2	&1992.99	&11.59	&0.84	&11.00	&1.44  	&0.81	&19.9	    &0.66    &0.34    &69.8	   &M3   &I	&\\
J0334-2619	&LHS1558	&03 34 12.71	&$-$26 19 26.8	&1991.93	&12.80	&2.16	&11.87	&1.40  	&0.83	&35.9	    &0.60    &0.35    &117.4	   &M3   &I	&\\
J0340-3526$^{10}$&LP944-20	&03 39 34.98	&$-$35 25 46.4	&1991.70	&17.24	&2.93	&13.95	&3.25  	&1.25	&7.7	    &0.31    &0.31    &16.1 	   &M9   &I	&\\
J0344-4211	&LP995-74	&03 44 05.29	&$-$42 10 44.7 	&1991.78 	&12.76 	&1.86 	&11.35	&1.19 	&0.79 	&62.7/15.5  &$-$0.19 &$-$0.45 &145.6/36.0  &M1   &I/int	&80/20\\
J0347-2254	&LHS1592	&03 46 55.57	&$-$22 54 11.8	&1991.93	&14.08	&2.09	&12.54	&1.53  	&0.94	&30.2	    &$-$0.38 &$-$0.42 &81.6	   &M4   &I	&\\
J0353-3823	&LP944-70	&03 53 29.68	&$-$38 23 05.5	&1991.91	&12.49	&1.73	&11.46	&1.31  	&0.90	&44.6	    &0.12    &0.41    &90.9	   &M2.5 &I	&\\
J0402-4325$^{11}$ &WT133 	&04 02 13.84	&$-$43 25 22.9	&1992.99	&14.07	&2.18	&12.92	&1.57  	&0.89	&31.4	    &0.06    &$-$0.58 &86.3 	   &M4   &I	&\\
J0403-3754	&LHS1622	&04 03 29.94 	&$-$37 53 37.9 	&1991.91 	&13.14 	&2.07 	&11.94 	&1.42 	&0.91 	&33.0 	    &0.52    &0.49    &111.1	   &M3   &I	&\\
J0405-6259	&LHS5090	&04 04 31.58	&$-$62 59 10.4 	&1989.98 	&14.49 	&1.85 	&12.85 	&1.26 	&0.80 	&103.6/25.5 &0.25    &$-$0.46 &255.4/62.9  &M2.5 &I/int	&60/40\\
J0413-5352$^{12}$&LHS1639	&04 12 47.13 	&$-$53 52 08.5 	&1994.99 	&11.56 	&2.18 	&10.96  &1.36 	&0.98 	&29.3 	    &0.50    &0.60    &108.3	   &M3   &I	&\\
J0413-3729	&		&04 13 22.71	&$-$37 28 54.0 	&1992.01 	&13.05 	&2.61 	&12.36 	&1.11 	&0.89 	&34.0       &0.73    &$-$0.29 &127.3  	   &M0.5 &int	&\\
J0419-5714	&LHS1656	&04 18 50.80	&$-$57 14 06.0	&1994.99	&11.02	&1.95	&10.75	&1.22  	&0.88	&42.5/10.7  &0.27    &0.75    &160.4/40.4  &M1   &I/int	&80/20\\
J0424-4551	&		&04 23 57.71 	&$-$45 50 38.8  &1992.99 	&16.25  &0.56   &16.00	&0.31  	&...  	&35.9	    &$-$0.10 &$-$0.51 &88.5	   &wd   &	&\\
J0433-3947$^{13}$&LHS1678	&04 32 42.33	&$-$39 47 05.3	&1992.08	&10.06	&2.31	&10.26	&1.15  	&0.83	&42.5/10.3  &0.27    &$-$0.97 &202.8/49.2  &M0.5 &I/int	&50/50\\
J0455-4238	&		&04 55 08.23	&$-$42 38 03.2 	&1990.07 	&12.94 	&1.97 	&12.01 	&1.10 	&0.73 	&109.5/30.3 &$-$0.20 &$-$0.29 &181.7/50.3  &M0.5 &I/int	&45/55\\
J0518-5256	&LP233-25	&05 17 30.73	&$-$52 56 16.2	&1991.12	&11.33	&2.27	&10.67	&1.30  	&0.86	&32.3	    &0.41    &0.22    &72.0	   &M2.5 &I	&\\
J0523-5609	&LHS5104	&05 22 57.74	&$-$56 09 01.0	&1991.12	&13.06	&2.10	&11.81	&1.42  	&0.93	&31.3	    &0.26    &0.47    &80.1	   &M3   &I	&\\
J0526-5501	&LP180-29	&05 26 28.00	&$-$55 01 01.9	&1991.12	&11.76	&2.04	&11.00	&1.38  	&0.87	&28.0	    &0.17    &0.46    &65.0	   &M3   &I	&\\
J0531-3012$^{14}$&LHS1767	&05 31 04.11	&$-$30 11 41.7	&1992.76	&11.49	&1.68	&10.68	&1.56  	&0.80	&11.7	    &0.35    &$-$0.45 &31.6	   &M4.5 &I	&\\
J0532-5657	&		&05 32 26.70	&$-$56 57 13.4	&1991.12	&13.33	&2.11	&12.14	&1.44  	&0.90	&33.5	    &0.04    &$-$0.42 &66.7	   &M3   &I	&\\
J0533-3952	&		&05 32 38.22	&$-$39 52 22.6 	&1992.02 	&14.06 	&1.63 	&12.77 	&1.26 	&0.78 	&96.7/23.8  &$-$0.10 &0.33    &155.8/38.4  &M2.5 &I/int	&90/10\\
J0540-4011	&L378-6		&05 40 02.63 	&$-$40 11 17.3 	&1992.02 	&11.36 	&1.66 	&10.43  &1.21 	&0.91 	&38.4 	    &$-$0.16 &0.34    &69.2	   &M1   &I	&\\
J0541-5349	&		&05 41 27.18	&$-$53 49 17.9 	&1991.12 	&11.82 	&1.79 	&11.77 	&1.13 	&0.82 	&90.3/23.6  &0.10    &0.37    &162.6/42.5  &M0.5 &I/int	&65/35\\
J0542-3618	&WT2431		&05 41 52.09	&$-$36 17 52.5 	&1992.00 	&12.15 	&1.66 	&11.77 	&1.02 	&0.74 	&38.7       &0.28    &$-$0.31 &77.0        &M0   &int	&\\
J0544-4108$^{15}$&		&05 43 46.37	&$-$41 08 04.7	&1992.02	&12.68	&1.76	&11.21	&1.53  	&0.87	&16.5	    &0.18    &$-$0.59 &48.5	   &M4   &I	&\\
J0552-5507	&LHS1793	&05 52 29.31	&$-$55 06 39.6 	&1993.89 	&11.18 	&2.22 	&10.36 	&1.11 	&0.83 	&50.1/13.6  &$-$0.27 &$-$0.61 &159.1/43.2  &M0.5 &I/int	&75/25\\
J0604-5519$^{16}$&LHS1815	&06 04 19.81 	&$-$55 18 49.3 	&1993.89 	&10.50 	&2.11 	&10.07  &1.28 	&0.78 	&26.4 	    &0.69    &0.36    &97.6	   &M2.5 &I	&\\
J0605-3742	&LP949-17	&06 04 51.30	&$-$37 41 42.0	&1991.93	&12.38	&2.02	&11.59	&1.34  	&0.82	&42.2	    &0.11    &0.46    &94.0	   &M2.5 &I	&\\
J0610-3740	&		&06 09 39.83	&$-$37 39 33.3	&1991.93	&18.62	&2.17	&15.35	&2.14  	&0.91	&41.9	    &0.11    &0.35    &73.5	   &M5.5 &I	&\\
J0611-4324$^{17}$&LHS1831	&06 10 52.72	&$-$43 24 24.2	&1991.87	&10.67	&2.00	&9.58	&1.34  	&0.90	&16.3	    &0.14    &0.76    &59.5	   &M2.5 &I	&\\
J0619-5124	&		&06 18 51.22	&$-$51 24 27.1	&1993.96	&14.07	&1.87	&13.03	&1.47  	&0.83	&45.1	    &0.16    &$-$0.29 &70.5	   &M3.5 &I	&\\
J0620-4144	&		&06 19 58.62 	&$-$41 43 35.8  &1991.93   	&15.38 	&$-$0.42&15.09	&$-$0.05&...	&39.7	    &0.19    &$-$0.38 &79.0	   &wd   &	&\\
J0629-5151	&		&06 28 44.22	&$-$51 50 55.5	&1992.09	&17.94	&2.25	&15.28	&2.01  	&0.97	&47.2	    &$-$0.12 &0.33    &78.3	   &M5.5 &I	&\\
J0653-5307	&LHS1877	&06 53 19.36	&$-$53 07 14.2	&1993.21	&10.18	&2.15	&10.33	&1.15  	&0.92	&44.1	    &$-$0.07 &0.48    &102.4	   &M0.5 &I	&\\
J0710-5704$^{18}$&		&07 09 37.29	&$-$57 03 45.4  &1992.17   	&11.93  &2.11   &10.66  &1.48  	&0.85	&15.1  	    &0.33    &0.31    &32.2	   &M3.5 &I	&\\
\hline\\
\multicolumn{14}{l}{$9$ distance is 17 pc from the Yale parallax catalogue, 22 pc from the ARICNS}\\
\multicolumn{14}{l}{$10$ already known brown dwarf at 5 pc \citep{tinney1996}}\\
\multicolumn{14}{l}{$11$ already identified by \cite{henry2002} at 30.2 pc}\\
\multicolumn{14}{l}{$12$ distance is 23 pc from the ARICNS}\\
\multicolumn{14}{l}{$13$ distance is 30 pc from the ARICNS}\\
\multicolumn{14}{l}{$14$ distance is 19 pc from the ARICNS}\\
\multicolumn{14}{l}{$15$ already identified by \cite{scholz2002} at 20 pc}\\
\multicolumn{14}{l}{$16$ distance is 26 pc from the Yale parallax catalogue, 29 pc from the Hipparcos catalogue}\\
\multicolumn{14}{l}{$17$ distance is 12 pc from the ARICNS}\\
\multicolumn{14}{l}{$18$ already identified by \cite{scholz2002} at 13 pc}\\
\end{tabular}
}}
\end{table*}

\begin{table*}
\rotatebox{90}{
{\scriptsize 
\begin{tabular}{llcccccrccrrrrlll}
\multicolumn{14}{l}{{\bf \normalsize Table 1.} \normalsize Continued\vspace*{0.3cm}}\\
\hline
name	&other  &$\alpha_{J2000}$&$\delta_{J2000}$&UKST		&$R$	&$B_J-R$ &\multicolumn{1}{c}{$I$}	&$I-J$	&$J-K_s$	
&\multicolumn{1}{c}{$d$}	&\multicolumn{1}{c}{$\mu_x$}	&\multicolumn{1}{c}{$\mu_y$}	
&\multicolumn{1}{c}{$V_t$}	&type &\multicolumn{2}{l}{population \%} \\	
APMPM &name	&		&		&epoch		&UKST	&UKST	&DENIS	&DENIS	&DENIS	&\multicolumn{1}{c}{(pc)}
&\multicolumn{2}{c}{($''$ yr$^{-1}$)}	&(km s$^{-1}$)	& && \\
\hline
J1039-2602	&LHS2294	&10 39 13.89	&$-$26 01 38.8 	&1993.08 	&16.19 	&1.89 	&13.75 	&1.44 	&0.77 	&68.9/25.9  &$-$0.56 &0.24    &199.2/74.9  &M3   &I/int	&55/45\\
J1042-3112	&LP904-51	&10 41 43.77	&$-$31 11 52.7	&1993.15	&15.10	&1.70	&12.84	&1.84  	&0.88	&19.3	    &0.37    &$-$0.25 &41.2        &M5   &I	&\\
J1051-2413	&LP849-16	&10 50 59.70	&$-$24 12 44.4 	&1993.08 	&14.61 	&1.67 	&12.63 	&1.30 	&0.75 	&77.4/20.2  &$-$0.18 &0.38    &154.1/40.2  &M2.5 &I/int	&40/60\\
J1106-2031	&LP791-26	&11 05 59.67	&$-$20 31 22.8	&1992.25	&15.82	&1.95	&13.83	&1.62   &0.86	&43.2	    &$-$0.37 &0.11    &79.9        &M4   &I	&\\
J1114-2529	&LHS 2376	&11 13 42.33	&$-$25 29 16.2	&1990.32	&14.60	&2.07	&13.00	&1.55   &0.89	&34.7	    &$-$0.45 &0.07    &74.8        &M4   &I	&\\
J1117-2757$^{19}$&LHS2385	&11 16 37.90	&$-$27 57 12.6	&1991.20	&11.80	&1.96	&10.84	&1.42  	&0.86	&20.2	    &$-$0.52 &$-$0.83 &93.8        &M3   &I	&\\
J1120-3056	&LP906-15	&11 20 15.99	&$-$30 56 09.0 	&1991.20 	&16.54 	&1.99 	&14.92 	&1.49 	&0.61 	&103.1/41.8 &$-$0.39 &0.17    &210.1/85.2  &M3.5 &I/int	&55/45\\
J1120-2230	&LHS2396	&11 20 18.23	&$-$22 30 05.8	&1992.25	&16.68	&2.47	&14.39	&2.05  	&0.84	&29.8	    &$-$0.43 &0.28    &72.0        &M5.5 &I	&\\
J1120-2802	&LP906-14	&11 20 20.36	&$-$28 01 56.9  &1991.20  	&13.88  &2.26 	&13.40	&1.01 	&0.80 	&86.3/43.4  &$-$0.48 &$-$0.05 &197.4/99.3  &M0   &int/II&75/25\\
J1218-2902	&LHS323		&12 17 30.67	&$-$29 02 20.1 	&1993.39 	&15.76 	&2.12 	&13.96 	&1.45 	&0.88 	&28.3       &$-$1.10 &$-$0.05 &147.6 	   &M3   &int	&\\
J1223-3010	&LP908-48	&12 22 37.20	&$-$30 09 30.3	&1993.39	&14.78	&2.01	&13.17	&1.51  	&0.79	&42.2	    &$-$0.31 &$-$0.02 &62.0        &M3.5 &I	&\\
J1224-2758$^{20}$&LHS325a	&12 23 56.73	&$-$27 57 48.5	&1993.39	&17.15	&2.37	&14.19	&2.30  	&0.88	&20.2	    &$-$1.25 &0.33    &123.5       &M6.5 &I	&\\
J1226-3121	&LP908-61	&12 25 52.57	&$-$31 20 48.3 	&1993.39 	&15.00 	&2.14 	&13.60 	&1.29 	&... 	&125.7/32.5 &$-$0.31 &0.03    &184.7/47.8  &M2.5 &I/int	&80/20\\
J1228-2746	&LHS2561	&12 27 35.03	&$-$27 46 16.2	&1993.39	&14.22	&2.20	&12.84	&1.54   &0.71	&33.4	    &$-$0.49 &0.31    &91.8        &M4   &I	&\\
J1230-2824	&LP909-6	&12 30 19.65	&$-$28 24 28.8	&1993.39	&13.70	&2.60	&12.40	&1.48  	&0.97	&33.1	    &$-$0.29 &$-$0.26 &61.2        &M3.5 &I	&\\
J1316-3229	&LHS2706	&13 15 34.29	&$-$32 28 34.7 	&1991.21 	&14.07 	&1.99 	&13.06 	&1.28 	&0.90 	&103.6/26.3 &$-$0.45 &$-$0.24 &250.4/63.6  &M2.5 &I/int	&70/30\\
J1344-2902	&LP911-47	&13 44 23.77	&$-$29 01 44.7 	&1992.19 	&14.94 	&2.53 	&13.72 	&1.26 	&0.79 	&153.3/37.8 &$-$0.31 &$-$0.02 &225.3/55.5  &M2.5 &I/int	&30/70\\
J1357-2828 	&LP912-35	&13 57 21.61	&$-$28 27 48.8	&1992.19	&12.83	&1.63	&11.88	&1.59  	&1.12	&18.7	    &$-$0.33 &0.03    &29.3        &M4   &I	&\\
J1407-3018$^{21}$ &LHS2859	&14 06 49.76	&$-$30 18 27.0	&1992.20	&15.65	&2.43	&13.38	&2.09   &1.04	&17.9	    &$-$0.82 &0.00    &69.6        &M5.5 &I	&\\
J1954-4748$^{22}$&LHS480	&19 54 00.15	&$-$47 48 27.8	&1990.73	&11.20	&1.82	&10.22	&1.13  	&0.77	&44.2/11.5  &$-$0.09 &$-$1.02 &214.5/55.8  &M0.5 &I	&50/50\\
J1957-4216	&		&19 56 57.35	&$-$42 16 17.2	&1993.62	&16.99	&2.27	&14.34	&2.07  	&0.61	&28.6	    &0.11    &$-$1.02 &139.6       &M5.5 &I	&\\
J1957-3251	&		&19 56 59.43	&$-$32 50 47.0	&1991.54	&14.79	&1.94	&12.73	&1.61  	&0.99	&26.6	    &$-$0.11 &$-$0.32 &42.9        &M4   &I	&\\
J2004-3142$^{23}$&LHS3516	&20 04 06.45	&$-$31 41 42.3	&1991.54	&13.40	&1.30	&11.44	&1.47  	&0.89	&21.8	    &0.34    &$-$0.77 &86.8        &M3.5 &I	&\\
J2007-3842	&		&20 07 10.64	&$-$38 42 08.4  &1993.62  	&15.75  &1.77  	&14.90 	&1.17	&0.78 	&84.3/42.5  &$-$0.11 &$-$0.50 &204.6/103.1 &M1   &int/II&50/50\\
J2008-5244	&LHS3525	&20 08 07.28 	&$-$52 44 21.6 	&1990.73 	&12.25 	&1.84 	&10.74  &1.20 	&0.91 	&46.1 	    &0.14    &$-$0.46 &104.9       &M1   &I	&\\
J2024-5102	&LHS3546	&20 23 53.38 	&$-$51 02 02.3 	&1991.59 	&12.04 	&1.93 	&11.53 	&1.07 	&0.63 	&26.9       &0.42    &$-$0.46 &79.1        &M0.5 &int	&\\
J2045-4218	&L423-44	&20 45 03.40 	&$-$42 17 41.3 	&1991.68 	&10.42 	&1.81 	&10.32 	&0.90 	&0.91 	&71.3/39.5  &0.38    &0.02    &128.4/71.1  &K7   &I/int	&40/60\\
J2101-4125	&		&21 01 03.44	&$-$41 14 31.8	&1991.68	&13.46	&1.94	&11.50	&1.55  	&0.90	&17.5	    &0.36    &$-$0.25 &36.5        &M4   &I	&\\
J2101-4907	&		&21 01 07.50	&$-$49 07 23.7	&1992.57	&11.25	&1.76	&10.52	&1.43  	&0.92	&16.5	    &$-$0.31 &$-$0.15 &26.6        &M3   &I	&\\
J2103-3022	&LHS3616	&21 03 17.02 	&$-$30 22 21.3 	&1989.76 	&12.83 	&1.89 	&11.29 	&1.21 	&0.85 	&57.4/14.1  &0.00    &$-$0.53 &144.2/35.4  &M1   &I/int	&80/20\\
J2103-5023$^{24}$&LHS3615	&21 03 21.30	&$-$50 22 50.1	&1992.57	&11.22	&1.38	&10.48	&1.44  	&0.86	&15.5	    &0.30    &$-$0.35 &33.8        &M3   &I	&\\
J2104-3828$^{25}$&		&21 03 30.48	&$-$38 27 47.7  &1991.68  	&14.39  &1.65  	&13.65 	&1.06 	&0.67 	&76.2/39.4  &$-$0.40 &$-$0.41 &206.9/107.0 &M0.5 &int/II&75/25\\
J2105-5235	&		&21 04 41.95 	&$-$52 34 40.7 	&1992.57 	&17.04 	&2.51 	&14.80  &2.12 	&0.99 	&33.1 	    &0.20    &$-$0.27 &53.3        &M5.5 &I	&\\
J2106-5143	&LP280-56	&21 06 26.95	&$-$51 42 36.7	&1992.57	&10.33	&1.63	&10.19	&1.26   &0.86	&30.1	    &$-$0.36 &$-$0.21 &59.9        &M2.5 &I	&\\
J2109-4004$^{26}$&		&21 08 30.86 	&$-$40 03 45.8	&1991.76 	&13.21 	&1.78 	&11.30  &1.35 	&0.83 	&35.8 	    &0.44    &$-$0.39 &100.1       &M2.5 &I	&\\ 
J2119-2817	&LP929-26	&21 19 11.96 	&$-$28 17 22.6 	&1989.76 	&14.21 	&2.08 	&12.61 	&1.25 	&0.80 	&95.1/23.4  &0.01    &$-$0.41 &184.8/45.5  &M1   &I/int	&80/20\\
J2134-4316$^{27}$&WT792		&21 34 22.10	&$-$43 16 06.0	&1993.61	&14.83	&2.50	&12.78	&2.01  	&1.07	&14.9	    &0.14    &$-$0.75 &53.7        &M5.5 &I	&\\
J2150-2731	&LHS5373	&21 50 23.33	&$-$27 31 15.8	&1995.63	&15.13	&2.35	&13.12	&1.51  	&0.93	&41.8	    &0.47    &$-$0.15 &97.1        &M3.5 &I	&\\
J2153-3121	&LP930-59	&21 53 02.73	&$-$31 20 49.4	&1995.63	&16.82	&2.16	&14.26	&1.76  	&0.89	&41.6	    &0.37    &$-$0.16 &78.9        &M5   &I	&\\
J2159-3126$^{28}$&LHS3738	&21 58 49.20	&$-$32 26 24.8	&1995.63	&14.87	&2.09	&12.44	&1.68  	&1.00	&20.2	    &$-$0.41 &$-$0.32 &49.8        &M4.5 &I	&\\
J2159-4612	&		&21 59 22.02 	&$-$46 11 57.0 	&1993.61 	&15.61 	&2.10 	&14.40 	&1.45 	&0.73 	&92.1/34.7  &0.42    &$-$0.13 &192.1/72.4  &M3   &I/int	&80/20\\
\hline\\
\multicolumn{14}{l}{$19$ distance is 16 pc from the ARICNS}\\
\multicolumn{14}{l}{$20$ already identified by \cite{phanbao2001} at 22.9 pc}\\
\multicolumn{14}{l}{$21$ already identified by \cite{phanbao2001} at 20 pc}\\
\multicolumn{14}{l}{$22$ distance is 25 pc from the Yale parallax catalogue}\\
\multicolumn{14}{l}{$23$ distance is 21 pc from the ARICNS}\\
\multicolumn{14}{l}{$24$ distance is 17 pc from the ARICNS}\\
\multicolumn{14}{l}{$25$ already identified by \cite{scholz2002} as a sdK5 at 105 pc}\\
\multicolumn{14}{l}{$26$ already identified by \cite{scholz2002} at 20 pc}\\
\multicolumn{14}{l}{$27$ already identified by \cite{phanbao2001} at 17.5 pc}\\
\multicolumn{14}{l}{$28$ distance is 24 pc from the ARICNS}\\
\end{tabular}
}}
\end{table*}

\begin{table*}
\rotatebox{90}{
{\scriptsize 
\begin{tabular}{llcccccrccrrrrlll}
\multicolumn{14}{l}{{\bf \normalsize Table 1.} \normalsize Continued\vspace*{0.3cm}}\\
\hline
name	&other  &$\alpha_{J2000}$&$\delta_{J2000}$&UKST		&$R$	&$B_J-R$ &\multicolumn{1}{c}{$I$}	&$I-J$	&$J-K_s$	
&\multicolumn{1}{c}{$d$}	&\multicolumn{1}{c}{$\mu_x$}	&\multicolumn{1}{c}{$\mu_y$}	
&\multicolumn{1}{c}{$V_t$}	&type &\multicolumn{2}{l}{population \%} \\	
APMPM &name	&		&		&epoch		&UKST	&UKST	&DENIS	&DENIS	&DENIS	&\multicolumn{1}{c}{(pc)}
&\multicolumn{2}{c}{($''$ yr$^{-1}$)}	&(km s$^{-1}$)	& && \\
\hline
J2202-3705$^{29}$&LHS3746	&22 02 28.65	&$-$37 04 50.3 	&1990.78 	&10.44 	&2.36 	&9.07 	&1.39 	&0.82 	&10.4 	    &0.78    &$-$0.17 &39.4        &M3   &I	&\\
J2205-3127	&LHS3753	&22 04 37.77 	&$-$31 27 12.2  &1995.63   	&16.73  &1.38	&16.12	&0.42	&...	&31.7	    &0.30    &$-$0.40 &75.1        &wd   &	&\\  
J2221-4218$^{30}$&LHS3798	&22 20 51.09 	&$-$42 18 23.0 	&1988.45 	&11.68 	&2.11 	&10.84 	&0.95 	&0.80 	&38.3 	    &0.52    &$-$0.29 &108.9       &M0   &int	&\\
J2222-4209	&LHS5384	&22 21 49.74	&$-$42 08 59.2	&1990.78	&13.36	&1.99	&11.79	&1.53  	&0.94	&21.0	    &$-$0.19 &$-$0.49 &52.8        &M4   &I	&\\
J2223-4328	&LHS3800	&22 23 08.90  	&$-$43 27 35.3 	&1990.78 	&13.85 	&1.54 	&12.23 	&1.33 	&0.75 	&15.7       &0.79    &$-$0.37 &64.7        &M2.5 &int	&\\
J2231-2800	&LP932-34	&22 31 04.83	&$-$27 59 51.2	&1995.78	&15.12	&2.17	&13.07	&1.57  	&0.69	&34.1	    &$-$0.03 &$-$0.47 &76.0        &M4   &I	&\\
J2231-2752	&LP932-36	&22 31 20.44  	&$-$27 51 35.8 	&1995.78 	&16.18 	&2.00 	&14.14 	&1.34 	&0.81 	&133.1/37.1 &0.32    &$-$0.13 &220.8/61.5  &M2.5 &I/int	&65/35\\
J2248-3042	&LP932-77	&22 48 06.01  	&$-$30 42 10.9 	&1995.78 	&12.11 	&1.48 	&11.47 	&0.98 	&0.83 	&110.0/41.2 &0.35    &$-$0.15 &198.1/74.2  &M0   &I/int	&20/80\\
J2250-2952	&LP932-7	&22 50 16.54	&$-$29 52 13.5	&1991.75	&14.98	&2.22	&12.90	&1.70  	&0.79	&24.2	    &$-$0.29 &$-$0.37 &53.9        &M4.5 &I	&\\
J2308-2754$^{31}$&LHS3898	&23 07 46.23 	&$-$27 54 22.8 	&1991.75 	&11.10 	&1.78 	&10.00  &1.17 	&0.89 	&35.5 	    &0.64    &$-$0.09 &109.4       &M1   &I	&\\
J2308-4644	&		&23 07 57.20  	&$-$46 44 00.1 	&1991.70 	&15.36 	&2.37 	&13.48 	&1.25 	&0.92   &34.6 	    &0.59    &0.12    &98.4        &M1   &int	&\\
J2316-3151	&LP933-4	&23 16 24.06	&$-$31 50 31.6	&1996.69	&14.03	&2.08	&12.01	&1.41  	&0.83	&36.7	    &0.17    &$-$0.26 &53.9        &M3   &I	&\\
J2318-4608	&LHS3924	&23 17 32.42 	&$-$46 08 10.8 	&1991.70 	&13.16 	&1.66 	&11.54  &1.42 	&0.74 	&27.2 	    &0.51    &$-$0.18 &69.6        &M3   &I	&\\ 
J2318-4819$^{32}$&LHS3925	&23 17 50.02 	&$-$48 18 42.4 	&1992.87 	&12.62 	&2.04 	&10.88  &1.39 	&0.80 	&24.5 	    &0.28    &$-$0.69 &85.9        &M3   &I	&\\ 
J2318-3027	&LHS3927	&23 18 15.36	&$-$30 27 10.6	&1996.69	&13.53  &1.73	&12.67	&0.96	&0.74	&38.9	    &0.42    &$-$0.83 &175.5       &M0   &II	&\\
J2318-2805	&LP933-6	&23 18 25.15  	&$-$28 05 11.5 	&1996.69 	&16.70 	&1.86 	&14.33 	&1.45 	&0.83 	&88.9/33.6  &0.33    &$-$0.13 &147.5/55.7  &M3   &I/int	&90/10\\
J2327-4651$^{33}$&LHS3949	&23 26 37.44 	&$-$46 51 03.1 	&1991.70 	&14.81 	&1.97 	&12.49  &1.62 	&0.80 	&23.2 	    &0.41    &$-$0.35 &59.4        &M4   &I	&\\
J2331-2750$^{34}$&		&23 31 21.61	&$-$27 49 52.7	&1996.69	&18.06	&2.42	&14.25	&2.59  	&1.04	&15.8  	    &0.07    &0.77    &57.7        &M7   &I	&\\
J2338-6906$^{35}$&LHS3988	&23 38 10.33  	&$-$69 05 56.8 	&1987.73 	&11.67 	&1.92 	&10.94 	&1.17 	&0.81 	&54.7/13.6  &0.89    &$-$0.12 &233.4/58.0  &M1   &I/int	&90/10\\
J2354-3316$^{36}$&LHS4039+4040	&23 54 01.06	&$-$33 16 25.6	&1996.61	&10.89	&1.17	&10.84	&1.38  	&0.89	&25.9	    &$-$0.32 &$-$0.33 &56.5        &M3   &I	&\\
J2355-2813	&LP935-40	&23 54 44.54	&$-$28 12 41.7	&1990.80	&17.30	&1.89	&14.62	&1.80  	&0.73	&45.8	    &0.35    &$-$0.12 &80.3        &M5   &I	&\\
J2358-3407	&LP987-60	&23 57 54.95	&$-$34 07 18.7	&1996.61	&17.61	&2.45	&15.03	&2.33  	&0.96	&28.9	    &0.13    &$-$0.35 &50.7        &M6.5 &I	&\\
J2359-6246$^{37}$&		&23 58 42.29	&$-$62 45 43.6	&1993.64	&14.91	&2.48	&13.29	&1.87  	&0.93	&22.6	    &0.60    &0.13    &65.3        &M5.5 &I	&\\
\hline\\
\multicolumn{14}{l}{$29$ distance is 10 pc from the ARICNS}\\
\multicolumn{14}{l}{$30$ distance is 51 pc from the Hipparcos catalogue}\\
\multicolumn{14}{l}{$31$ distance is 45 pc from the Yale parallax catalogue, 28 pc from the ARICNS}\\
\multicolumn{14}{l}{$32$ distance is 22 pc from the ARICNS}\\
\multicolumn{14}{l}{$33$ distance is 24 pc from the ARICNS}\\
\multicolumn{14}{l}{$34$ already identified by \cite{phanbao2001} at 17.3 pc}\\
\multicolumn{14}{l}{$35$ distance is 35 pc from the ARICNS}\\
\multicolumn{14}{l}{$36$ distance is 21 pc from the ARICNS}\\
\multicolumn{14}{l}{$37$ already identified by \cite{scholz2002} at 11 pc}\\
\end{tabular}
}}
\end{table*}

\section{Conclusion}
Thanks to the deep and large surveys at optical and near-infrared wavelengths, many 
low-luminosity nearby stars are identified. Most of the nearest stars turned out to have a large
proper motion. In this paper, we present the identification of 107 stars closer than 50~pc among
high proper motion stars cross-identified with DENIS. Their large proper motions exclude that they 
are giants. They are mainly M-stars, while 4 
stars are white dwarfs. 31 stars among them have already measured distances. In addition, 40 
stars may enter in the 50 pc limit depending on which 
population they belong to. 6 stars among them have previously measured distances.

15 new stars are within the 25~pc limit 
of the CNS3 catalogue. 5 stars, LHS5045, LP225-57, LP831-45, LHS1767, and WT792, are closer than 
15~pc,with 1 object, LP225-57, at 9.5~pc. In order to 
insure the spectral types and therefore distances, we plan to obtain low-resolution spectroscopy of the
closest candidates.

\begin{acknowledgements}
The authors thank Guy Simon who provided photometric measures in a strip not available in
the PDAC database, the whole DENIS 
staff and all the DENIS observers who collected the data. The DENIS project is supported 
by the SCIENCE and the Human Capital and Mobility plans of the European Commission under 
grants CT920791 and CT940627 in France, by l'Institut National des Sciences de l'Univers, 
the Minist\`ere de l'\'education Nationale and the Centre National de la Recherche 
Scientifique (CNRS) in France, by the State of Baden-W\"urtemberg in Germany, by the 
DGICYT in Spain, by the Sterrewacht Leiden in Holland, by the Consiglio Nazionale delle
Ricerche (CNR) in Italy, by the Fonds zur F\"orderung der wissenschaftlichen Forschung 
and Bundesministerium f\"ur Wissenschaft und Forschung in Austria, and by the ESO C \& EE 
grant A-04-046.
\end{acknowledgements}

\end{document}